\documentclass[copyright,creativecommons]{eptcs}
\providecommand{\event}{SR 2013} 
\usepackage{breakurl}             
\usepackage{listings}
\usepackage{hyperref}
\usepackage{graphicx}
\usepackage{float}
\usepackage{algorithmicx}
\usepackage{algorithm}
\usepackage{algpseudocode}
\usepackage{amsmath,amssymb}
\usepackage{soul}
\usepackage{color}
\usepackage{tikz}
\usepackage{wrapfig}
\usepackage{multirow}
\usepackage{subfig}
\usepackage{times}

\definecolor{green}{rgb}{0,.7,0}

\newcommand{\sectref}[1]{Section~\ref{#1}}
\newcommand{\figref}[1]{Figure~\ref{#1}}
\newcommand{\sectsectref}[2]{Sections~\ref{#1} and \ref{#2}}
\newcommand{\figfigref}[2]{Figures~\ref{#1} and \ref{#2}}

\newcommand{\trust}[1]{{\mathit{trust}_{red}}}
\newcommand{\trustij}{{\mathit{trust}_{ij}}}
\newcommand{\Cmin}{{C_\mathit{min}}}
\newcommand{\Cmax}{{C_\mathit{max}}}

\newcommand{\opF}{\mathsf{F}\,}

\newcommand{\coal}[1]{\lla #1 \rra}

\newcommand{\lla}{\langle\!\langle}
\newcommand{\rra}{\rangle\!\rangle}

\newcommand{\probop}{\mathsf{P}}
\newcommand{\prob}[1]{\probop_{#1}} 
\newcommand{\rewardop}{\mathsf{R}}
\newcommand{\reward}[2]{\rewardop^{#2}_{#1}} 

\tikzstyle{state1}=[draw, rectangle, inner sep=0mm, minimum size=1mm]
\tikzstyle{state2}=[draw,diamond, inner sep=0mm, minimum size=1.5mm]
\tikzstyle{state3}=[draw,circle, inner sep=0mm, minimum size=1.2mm]

\newcommand{\startpara}[1]{{%
\vskip5pt\noindent
{\bf #1.}}}

\title{Strategic Analysis of Trust Models for User-Centric Networks}
\author{Marta Kwiatkowska
\institute{Department of Computer Science \\ University of Oxford} 
\and
David Parker
\institute{School of Computer Science \\ University of Birmingham} 
\and
Aistis Simaitis
\institute{Department of Computer Science \\ University of Oxford} 
}
\def\titlerunning{Strategic Analysis of Trust Models for User-Centric Networks}
\def\authorrunning{Kwiatkowska, Parker \& Simaitis}

\begin{document}
\maketitle

\begin{abstract}
We present a strategic analysis of a trust model that has
recently been proposed for promoting cooperative behaviour in user-centric networks.
The mechanism for cooperation is based on a combination of reputation and virtual currency schemes
in which service providers reward paying customers and punish non-paying ones
by adjusting their reputation, and hence the price they pay for services.
We model and analyse this system using PRISM-games,
a tool that performs automated verification and strategy synthesis
for stochastic multi-player games using the probabilistic alternating-time temporal logic with rewards (rPATL).
We construct optimal strategies for both service users and providers,
which expose potential risks of the cooperation mechanism
and which we use to devise improvements that counteract these risks.
\end{abstract}


\section{Introduction}

User-centric networks are designed to encourage users to act cooperatively,
sharing resources or services between themselves,
for example in order to provide connectivity in a mobile ad-hoc network.
The effectiveness of such networks is heavily dependent on their cooperation mechanisms,
which are often based on the use of incentives to behave unselfishly.
In this paper, we present an analysis of a cooperation mechanism for user-centric networks \cite{BPA+12},
which combines a \emph{reputation}-based incentive, used to establish a measure of \emph{trust} between users, 
and a \emph{virtual currency} mechanism used to ``buy'' and ``sell'' services.


The cooperation model proposed in \cite{BPA+12} was analysed formally by the authors using
probabilistic model checking \cite{AB12,AB12a}.
They verified several performance properties, specified in the probabilistic temporal logics PCTL and CSL,
on discrete- and continuous-time Markov chains models and,
in \cite{AB12}, also used Markov decision processes
to assess the worst-case performance of service providers.

In this paper, we take a different approach and study the cooperation mechanism
using \emph{strategy-based analysis}. The system is modelled as a
\emph{stochastic multi-player game}, in which service providers
and customers are modelled as players with objectives,
expressed in the logic probabilistic alternating-time temporal logic with rewards (rPATL)~\cite{CFK+13b}.
We model and analyse the cooperation mechanism using PRISM-games \cite{CFK+13},
a probabilistic model checker for stochastic multi-player games.
We use rPATL model checking to identify weaknesses in the cooperation mechanism
and then perform \emph{strategy synthesis} to discover important insights into the model:
firstly, we construct and visualise potential attacks or undesirable behaviour;
secondly, we develop improvements to the system that alleviate these problems
and check their correctness.



\startpara{Related work}
Game-theoretic techniques have been applied to a wide variety of problems in the context
of computer networks, from network security \cite{Roy10}
to self-organisation in ad-hoc networks \cite{FHB06}.
Of particular relevance to this paper is the work in \cite{LS12},
which gives a game-theoretic analysis of cooperative incentive schemes in mobile ad-hoc networks
and proposes the combination of trust and currency mechanisms used in \cite{BPA+12}.
Its effectiveness is analysed using a combination of theoretical and simulation results.
By contrast, we adopt a semi-automatic approach where the strategies are synthesised 
automatically by the tool from rPATL specifications, and are then analysed to understand and improve the cooperation mechanism.
The logic rPATL has been previously used to analyse cooperation incentives
in micro-grid energy management and decentralised agreement in sensor networks \cite{CFK+13b},
but a detailed strategy-based analysis was not performed.


\section{Modelling the Cooperation Mechanism}

\subsection{The cooperation mechanism}
\label{sec:mech}

The basic ideas behind the cooperation mechanism of~\cite{BPA+12} can be summarised as follows.
We assume a general model of \emph{providers} offering \emph{services} to \emph{requesters}.
Cooperation between \emph{users} of the network (requesters and providers)
is managed through a combination of reputation and virtual currency.

Reputation is captured by a discrete \emph{trust measure}, denoted $\trustij$,
representing the extent to which user $i$ trusts user $j$,
based on previous interactions between them and the recommendations provided by 
the other users in the network.
This is used to determine whether a service request from $j$ is accepted by $i$.
A \emph{trust level} $T_{ij}$ is computed as a weighted sum
$T_{ij}=\alpha\cdot\trustij+(1{-}\alpha)\cdot\mathit{recs}_{ij}$,
where $\mathit{recs}_{ij}$ is an ``indirect'' trust measure,
taken as the the average value of $\mathit{trust}_{kj}$ for other users $k$
(whereas $\trustij$ is called a ``direct'' measure of trust).
By default, $i$ will decide to accept $j$'s request if $T_{ij}$ is not below a pre-specified
\emph{service trust level}, denoted $\mathit{st_i}$.
The parameter $\alpha \in [0,1]$ controls the relative influence that the direct and
indirect measures of trust have on this decision.

The reputation scheme is then integrated with a virtual currency system,
where services are bought and sold between users,
and the cost paid to $i$ by $j$ for a service is a function of $\trustij$.
Assuming model parameters for minimum and maximum costs $\Cmin,\Cmax$ and threshold $T'$,
the cost is defined as 
\[C(\trustij)=\left\{\begin{array}{ll}
\Cmin+\frac{\Cmax-\Cmin}{T'}\cdot(T'-\trustij) & \mbox{if } \trustij<T' \\
\Cmin & \mbox{if } \trustij\geq T' \\
\end{array}\right.\]

Procurement of a service proceeds in several phases.
First, a requester $j$ chooses a provider $i$ and makes a request.
If $T_{ij}\geq st_i$, the request is accepted.
In this case, the two users then ``negotiate'' the service cost,
using the function of $\trustij$ given above.
The negotiation may, however, fail: with probability $c_i$, user $i$
cancels the accepted request; this represents the ``cooperative attitude''~\cite{AB12} of the provider $i$.
If not cancelled, the service is delivered and the requester chooses whether or not to pay the negotiated 
price to the provider. If payment is made, the provider increases the trust measure
of the requester by one unit. If not, the measure is decreased by $td_i$ units.
On encountering a requester for the first time, a provider shares the trust measure with
the other providers.
 

\subsection{A stochastic game model}

We build a model of the cooperation mechanism of~\cite{BPA+12}
as a (turn-based) stochastic multi-player game (SMG).
An SMG comprises a finite set of \emph{players} and a finite set of \emph{states}.
In each state, exactly one player chooses (possibly randomly)
from a set of available \emph{actions}. When an action is taken, the result
is a \emph{probabilistic transition}, i.e. a successor state is chosen according to a discrete probability distribution.
The choices for each player are made by a \emph{strategy},
which selects an action (or distribution over actions) based on the history of the SMG so far.
The strategies needed in this paper are \emph{memoryless} (i.e. history independent) and \emph{deterministic} (i.e. do not use randomisation).

We developed the SMG model using the PRISM-games model checker,
taking the PRISM model of~\cite{AB12} as a starting point.%
\footnote{All model/property files are available at: \url{http://www.prismmodelchecker.org/files/sr13trust/}}.
The SMG model has one player for each user in the network.
The choices made by a ``requester'' player model the decision of which provider
is selected at each point in the system execution.
In the basic model, the ``provider'' players do not have any choices to make;
later (in \sectsectref{sec:cost}{sec:incentives}),
we will add choices for these players in order to synthesise strategies that
can be used to improve the cooperation mechanism.
The stochastic aspects of the SMG model are primarily required to
model the fact that negotiations fail probabilistically.

We adopt the same basic network configuration as
used in the original analysis of the protocol~\cite{AB12}, which 
comprises 3 providers and 1 requester.
Even though this network is relatively small,
it still captures the fundamental aspects of the protocol.
For instance, observe that the decision whether to provide a service to a requester
does not depend on the trust level of other requesters in the network,
so incorporating more requesters does not offer any more information
about the dynamics of trust and provided services. 
On the other hand, as we will show, using three
service providers already allows us to identify
malicious strategies for the requester that can be generalised to 
an arbitrary number of providers (see, e.g., the discussion about
the number of unpaid requests in \sectref{sec:unpaid}).

The parameters of the cooperation mechanism are also taken from~\cite{AB12} 
and are as follows.
The trust measure is an integer in the range $0$ to $10$ and is initially $5$.
We use $\alpha{=}0.8$ to compute trust levels, unless stated otherwise,
and the service trust threshold $st_i$ is set to $5$ for all providers.
We use a negotiation failure probability of $c_i{=}0.05$ for all providers $i$,
and the parameters used to compute prices are fixed at $\Cmin{=}2,\Cmax{=}10$ and $T'{=}8$.


\section{A Strategy-based Analysis}

We now analyse the cooperation model described above,
showing how the interplay between the two key components of the protocol,
trust and virtual currency, affects the cooperation dynamics.
Our analysis is based on strategy synthesis for properties in the temporal logic rPATL \cite{CFK+13b}.
The logic combines features of the multi-agent logic ATL, the probabilistic logic PCTL,
and operators to reason about expected reward or cost measures.
A simple example of an rPATL formula is
$\coal{\{1,2\}} \prob{\geq 0.75} [\opF^{\leq5}\mathit{goal}]$,
which asks
``do players $1$ and $2$ have a (combined) strategy to ensure that the probability of reaching a
`$\mathit{goal}$' state within 5 steps is at least $0.75$,
regardless of the strategies of other players in the game?''.
Alternatively, we can use a numerical query such as
$\coal{\{1,2\}} \prob{\max=?} [\opF^{\leq5}\mathit{goal}]$:
``what is the maximum probability of reaching a
`$\mathit{goal}$' state within 5 steps that can be ensured by players 1 and 2?''.
An example of property to reason about rewards (or costs) is
$\coal{\{3\}} \reward{\leq 10}{r} [\opF^{\star}\mathit{goal}]$,
which asks ``does player $3$ have a strategy to ensure that the expected amount of reward
$r$ cumulated before reaching a `$\mathit{goal}$' state is at most 10?''.
The $\star$ parameter lets us specify what the total reward should be if a `$\mathit{goal}$' state
is \emph{not} reached: we can assign zero reward ($\star{=}0$), infinite reward ($\star{=}\infty$)
or allow reward to accumulate indefinitely ($\star{=}c$).
For precise details of the logic rPATL and its semantics, we refer the reader to \cite{CFK+13b}.



\subsection{Unpaid requests}\label{sec:unpaid}

\begin{figure}[htb]
\centering
\subfloat[\emph{Number} of unpaid services.]{
\label{fig:unpaid_orig}
\includegraphics[clip=true, trim=75 555 260 98, scale=0.55]{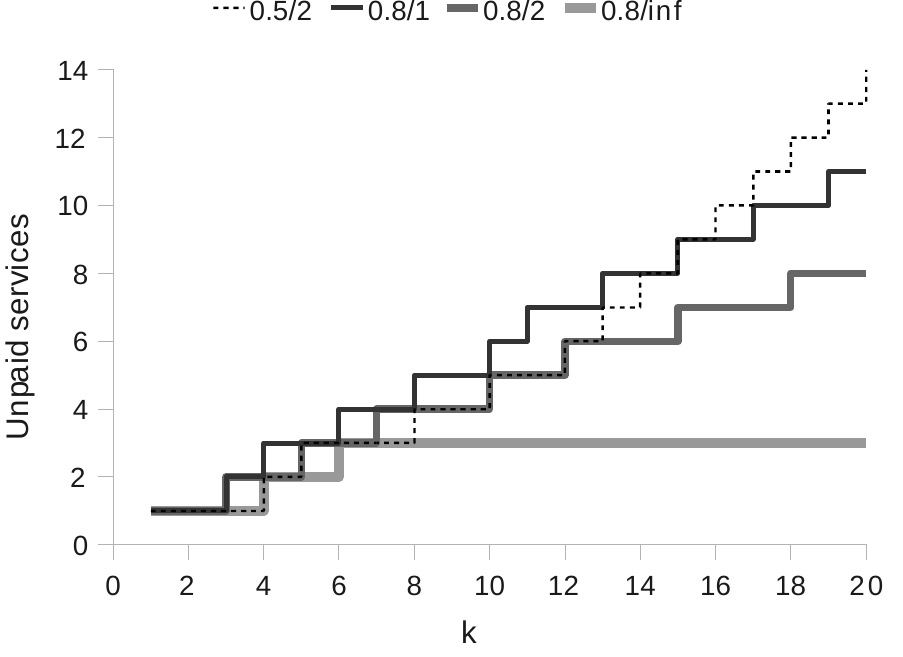}
}
\subfloat[\emph{Fraction} of unpaid services.]{
\label{fig:unpaid_perc}
\includegraphics[clip=true, trim=75 354 260 298, scale=0.55]{images/graph1}
}
\subfloat[Strategy for  0.5/2 and $k=13$.]{
\label{fig:unpaid_strat}
\includegraphics[clip=true, trim=75 139 240 490, scale=0.55]{images/graph1}
}
\caption{
Maximum unpaid services the requester can achieve in obtaining $k$ services.
}
\label{fig:unpaid_requests}
\end{figure}

First, we consider the extent to which the requester can obtain services without paying for them.
We analyse the maximum (expected) number of unpaid services that the requester can obtain
if its goal is to get $k$ services in total. This is expressed in rPATL as:
\[
\coal{\{requester\}} \reward{\max=?}{unpaid} [\opF^{c}\mathit{services\!=\!k}],
\]
where
$\mathit{unpaid}$ denotes a \emph{reward structure} assigning 1 to every unpaid request.
The results for various combinations of model parameters $\alpha$ and $td_i$ are shown
in \figref{fig:unpaid_requests} (we use $0.5$/$2$ to indicate that $\alpha=0.5$ and 
trust is decreased by $td_i=2$ units upon an unpaid service; $td_i=\mathrm{inf}$ means
that trust is reset to 0 upon an unpaid service).

\figfigref{fig:unpaid_orig}{fig:unpaid_perc}
show the number and fraction, respectively, of services that are unpaid, for a range of $k$.
From \figref{fig:unpaid_perc}, in particular,
we see that, for parameters $0.5$/$2$ and $0.8$/$\mathit{inf}$, the behaviour is fundamentally different from the other two
- the portion of requests converges to $1$ and $0$, respectively. For $0.8$/$\mathit{inf}$, this behaviour
is expected, because the trust measure is decreased to $0$ upon non-payment; however, the behaviour of $0.5$/$2$
represents an attack on the trust model allowing the requester to receive an unlimited
number of unpaid services for a fixed cost.
We synthesise an attacker (requester) strategy for our model with 3 providers,
for the case of acquiring $k=13$ services:
for a cost of 5 services, the requester can get an unlimited number of unpaid services.
We depict the strategy in \figref{fig:unpaid_strat}.
Arrows represent ``request-and-pay'' (white arrow)
and ``request-and-do-not-pay'' (grey arrow) actions of the optimal requester strategy,
depending on the number of services acquired so far.

This attack is possible if $st_i \le (1-\alpha)\cdot T_{\max}$ for some provider $i$,
where $T_{\max}$ is the maximum trust level among all providers.
We note that it is only viable if the network is sufficiently small since the fixed cost increases
with the number of providers sharing the trust information: to achieve the 
required indirect trust measure $\mathit{recs}_{ij}\ge \frac{st_i}{1-\alpha}$, the requester must
pay for a number of services proportional to the number of providers. However, in order to work, this
requires that all providers share their initial direct trust measure even though they have not encountered the
requester. 


\subsection{Cost of obtaining services}\label{sec:cost}

\begin{figure}[htb]
\centering
\subfloat[Minimum cost to obtain $k$ services.]{
\label{fig:min_cost}
\includegraphics[clip=true, trim=65 569 279 85, scale=0.55]{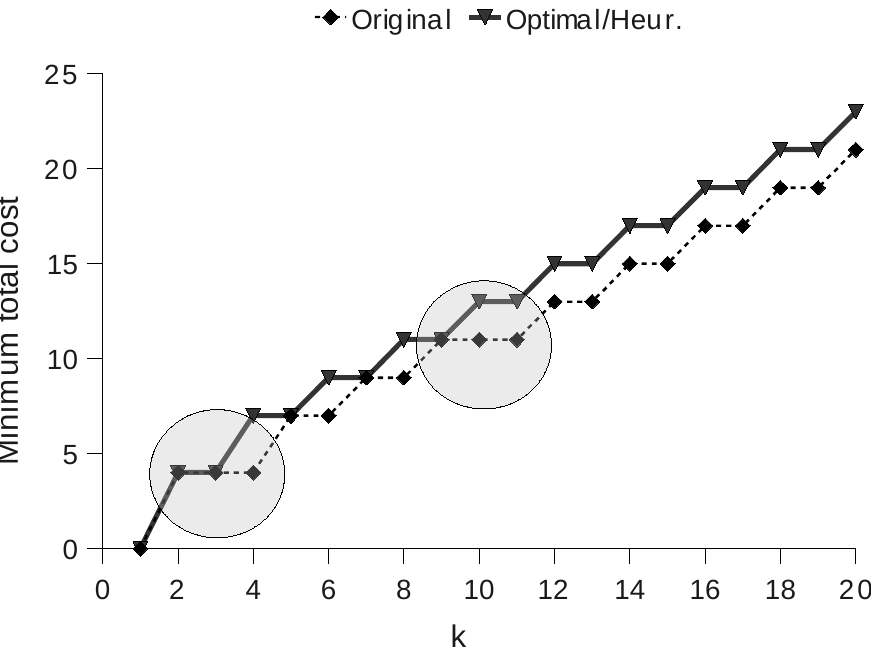}
}\ \ \ \ \ \ \ \ \ 
\subfloat[Example strategy for $k=13$.]{
\label{fig:cost_strat}
\includegraphics[clip=true, trim=67 350 269 275, scale=0.55]{images/graph2}
}
\caption{
Cost of $k$ services for requester and a strategy example.
}
\label{fig:currency}
\end{figure}

We now turn our attention to the virtual currency system, and study the minimum 
price at which the requester can buy $k$ services. For this, we use rPATL formula:
\[
\coal{\{requester\}} \reward{\min=?}{cost} [\opF^{\infty}\mathit{services\!=\!k}].
\]
Intuitively, the requester has a strategy to get one unpaid service for each
paid service by executing the following sequence: pay, not pay, pay, not pay, etc.  
However, a plot of the above property
(see highlighted sections of line `Original' in \figref{fig:min_cost}),
shows deviations from this pattern, where the requester can get
$4$ services for the price of $2$ and, similarly, $11$ services for the price of $9$.

We synthesise a strategy achieving this and depict it in \figref{fig:cost_strat}. We can see that
all paid requests are directed to one provider and the others only receive unpaid requests.
In fact, by exploiting the reputation system, the requester is even able to obtain $2$ unpaid requests from provider $2$.

Next, we devise a fix by changing the model to allow providers to manage the way
they share trust information between themselves: they can choose whether to share trust 
information after interaction with the requester. We synthesise the optimal
trust information sharing strategy for cooperating providers,
whose behaviour is shown as `Optimal/Heur.'\ in \figref{fig:min_cost}
and can be seen to avoid the above shortfall. Manual examination of the
synthesised strategy reveals a suitable heuristic whereby providers share trust information only
when its direct trust of the requester is smaller than that of the others.
We implement this heuristic in the model and find that
it yields the same model checking results as the optimal strategy.


\subsection{Provider selection incentives}\label{sec:incentives}

\begin{figure}[htb]
\centering
\subfloat[Original pricing scheme.]{
\label{fig:pricing_orig}
\includegraphics[clip=true, trim=65 545 268 97, scale=0.55]{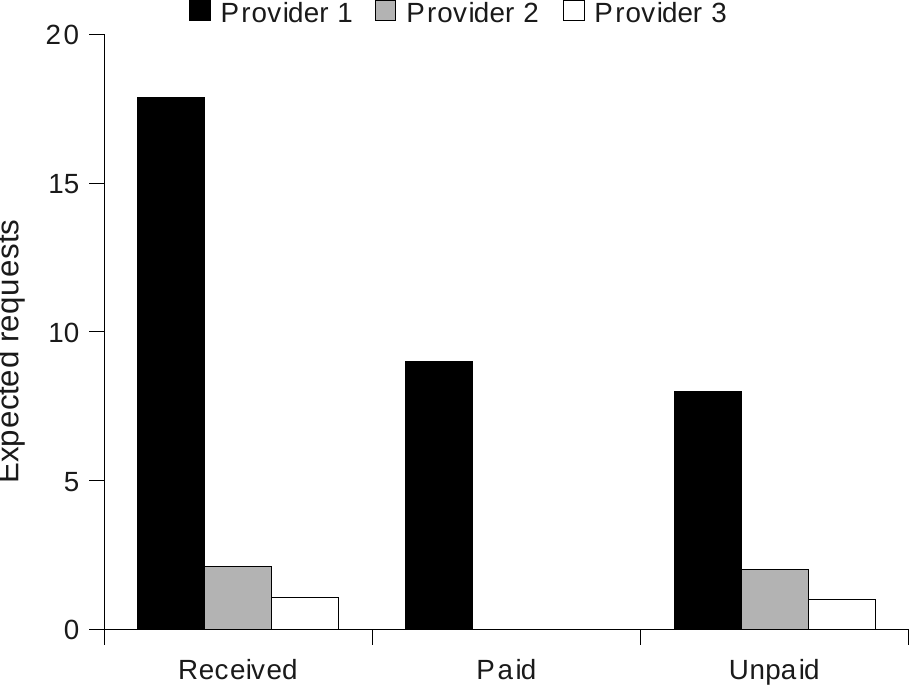}
}
\subfloat[Modified pricing scheme.]{
\label{fig:pricing_new}
\includegraphics[clip=true, trim=65 322 268 320, scale=0.55]{images/graph3}
}
\subfloat[Optimal strategy for 13 services.]{
\label{fig:pricing_strat}
\includegraphics[clip=true, trim=65 105 268 535, scale=0.55]{images/graph3}
}
\caption{
Distribution of requests among providers.
}
\label{fig:pricing}
\end{figure}

Another interesting feature revealed by the analysis of the strategy in the previous section
is that the proposed virtual currency system provides an incentive for the requester to
only ever pay for services from one provider (see \figref{fig:pricing_orig}). 
This is in
fact optimal behaviour because, in the computation of the service cost, only the direct
trust measure is used.  
This may or may not be a desired feature for the mechanism. We can show that a simple change that incorporates
the maximum difference between trust into the pricing model
(i.e., cost is now computed as $original\_cost+\max_{k}|\trustij-\mathit{trust}_{kj}|$,
where $original\_cost$ is the cost assigned by the pricing scheme from \sectref{sec:mech})
incentivises the requester to disperse its requests between service providers.

\figref{fig:pricing_new} shows the distribution of requests between providers and
\figref{fig:pricing_strat} depicts the actions of the optimal strategy in the new
pricing scheme.
This strategy contrasts with the strategy for the original mechanism from
\figref{fig:cost_strat}
because paid requests are now distributed uniformly across all the service providers.
This analysis of strategies has been performed using the ``strategy implementation'' feature
of PRISM-games, which allows the user to synthesise an optimal player strategy for some rPATL
formula, and then evaluate a second rPATL property on the modified SMG in which one coalition's
strategy is fixed using the previously synthesised one.
In this instance, we used the following rPATL formulae:
\[
\coal{\{requester\}} \reward{\min=?}{cost} [\opF^{\infty}\mathit{services\!=\!k}]\quad  
\text{and}\quad 
\coal{\emptyset} \reward{\min=?}{r} [\opF^{c}\mathit{services\!=\!k}],
\]
where the first formula was used to synthesise the strategy
 and the second formula is the one used to analyse it 
($r$ represents reward structures for Received, Paid, and Unpaid).


\section{Discussion and future work}

We have presented a strategy-based analysis of a cooperation mechanism for
user-centric networks, using automated verification of stochastic multiplayer games.
We have identified several undesirable properties of the model,
including attacks on the reputation system and inefficiencies of the virtual currency
mechanism. These would have been difficult to discover using conventional model checking.
Furthermore, we have shown that an analysis of optimal strategies for the model can
help us understand the incentives that the model introduces to the system 
and to devise and verify improvements. 

Our approach, which is based on probabilistic model checking,
builds and analyses a more detailed system model than
other game-theoretic analysis techniques, such as \cite{LS12}.
On the one hand, this may impose limitations on the scalability of our approach.
On the other hand, we are able to look at the protocol in fine detail and, as we have
shown in this paper, identify subtle problems that arise even with a small number of system components,
but which may also generalise to larger models.

There are many interesting directions for future work.
We plan to further develop our probabilistic model checker PRISM-games
to provide a wider range of analysis techniques.
For example, we plan to incorporate additional reward operators
dealing with limit averages and discounted sums.
We would also like to investigate extensions of our techniques
to incorporate partial-information strategies
or more complex solution concepts such as Nash and subgame-perfect equilibria.

\startpara{Acknowledgments}
The authors are part supported by ERC Advanced Grant VERIWARE,
the Institute for the Future of Computing at the Oxford Martin School
and EPSRC grant EP/F001096/1. 

\bibliographystyle{eptcs}
\bibliography{references}
\end{document}